\documentclass[a4paper,11pt]{article}
\usepackage{jinstpub} 
\usepackage{lineno}
\usepackage{pythonhighlight}
\usepackage{listings}
\usepackage{xcolor}
\usepackage{siunitx}

\title{Doberman: a modular and distributed slow control system for small- to medium-scale experiments}

\author[1]{Jaron~Grigat,\note{Corresponding author.}}
\author{Darryl~Masson,}
\author{Marc~Schumann}

\affiliation{Physikalisches Institut, Universit\"at Freiburg, D-79104 Freiburg, Germany}

\emailAdd{jaron.grigat@physik.uni-freiburg.de}

\abstract{
We present Doberman (Detector OBsERving and Monitoring ApplicatioN), a light- weight, modular, and open-source slow control system designed for small- to medium-scale physics experiments. Doberman addresses the gap between heavyweight industrial SCADA frameworks and ad hoc laboratory solutions by providing a flexible software architecture that supports heterogeneous instrumentation, distributed deployment, automated control, and robust alarm handling.

The web-based graphical user interface Doberview provides live and continuously updated visualization, configuration, and control of the entire experiment, supporting both routine operation and rapid response to exceptional conditions. 
Doberman has been deployed and validated in multiple experimental setups, ranging from a remotely operated underground gamma-ray spectrometer to a large, highly instrumented liquid xenon test facility with several hundred monitored quantities.

Doberman and Doberview are released under permissive open-source licenses, and the software, documentation, and example device integrations are publicly available.
}

\arxivnumber{2602.04580} 

\begin{document}
\maketitle
\flushbottom

\section{Introduction}
\label{sec:intro}

Many physics experiments rely on a slow control system to monitor and regulate auxiliary parameters such as voltages, currents, pressures, temperatures, rates, gas flows, or filling levels. In contrast to the main data-acquisition system, which records physics signals at high rates, a slow control system typically operates at comparatively low sampling frequencies of order \SI{1}{\hertz}. Despite this, its role is equally critical: it facilitates the operation of complex detectors, ensures stable detector conditions, protects sensitive equipment, provides automated alerts in case of anomalies, and supplies essential environmental and detector-state information used to correct and calibrate the physics data.

For large-scale experiments, these tasks are commonly handled by industrial Supervisory Control and Data Acquisition (SCADA) systems or by comprehensive open-source frameworks such as EPICS~\cite{epics,epics2}. While these solutions are powerful and highly reliable, they are often too complex, resource-intensive, or costly for small and medium-scale laboratory experiments and R\&D platforms. In practice, many such setups therefore operate either with ad-hoc scripts or without a unified slow control infrastructure at all, which can compromise long-term stability, data integrity, and operational safety.

Doberman (Detector OBsERving and Monitoring ApplicatioN) was originally developed to address this gap~\cite{doberman_og} and has since evolved substantially through nearly a decade of continuous use and development. It is a lightweight, modular, and open-source slow control system tailored to small and medium-scale experiments, while remaining scalable to setups with hundreds of monitored quantities. Its design is centered around a plugin-based architecture that allows a wide range of commercial and custom-built instruments to be integrated through device-specific drivers, while exposing a unified interface to the core system. This approach provides flexibility, simplifies maintenance, and enables the system to evolve as experimental setups change.

A key design goal of Doberman is operational robustness. The system is built to tolerate individual device failures, communication problems, and host outages without compromising the overall functionality of the slow control infrastructure. Measurements, configurations, and logs are stored in dedicated databases, enabling both long-term traceability and prompt, continuously updated visualization. In addition, Doberman includes a built-in automation and alarm framework that can execute control actions and trigger notifications in response to abnormal conditions, thereby reducing the need for continuous human supervision.

Equally critical is a responsive and intuitive graphical user interface. 
In practice, the personnel responsible for operating an experiment are not necessarily experts on the underlying slow control system, yet they must be able to interpret its output and react reliably in time-critical situations. 
A clear, well-structured user interface that presents relevant information at a glance and allows safe interaction with control elements is therefore essential for effective operation. 
Doberman addresses this requirement by providing a dedicated web-based interface that enables rapid assessment of detector status, straightforward configuration of the system, and controlled interaction with actuators, even under operational stress.

Here we describe the design and implementation of the Doberman slow control system. In 
\autoref{sec:doberman} we present the overall system architecture, including the distributed deployment model, the database backend, the communication and supervision infrastructure, and the abstraction of devices, sensors, and automated control logic. 
The graphical user interface, Doberview, which provides live visualization, configuration, and interaction with the system, is described in \autoref{sec:doberview}. 
Finally, \autoref{sec:deployments} summarizes several deployments of Doberman in running experiments, illustrating its scalability and suitability for long-term operation in heterogeneous laboratory environments.

\section{Doberman architecture}
\label{sec:doberman}

Doberman is designed to scale from compact single-host installations to distributed multi-host deployments. Its architecture combines a configuration database for system settings, a time-series database for slow control measurements, and a set of long-running processes (``Monitors'') that acquire data, execute automation logic, and distribute alarms. 

\begin{figure}[ht]
    \centering
    \includegraphics[width=\textwidth]{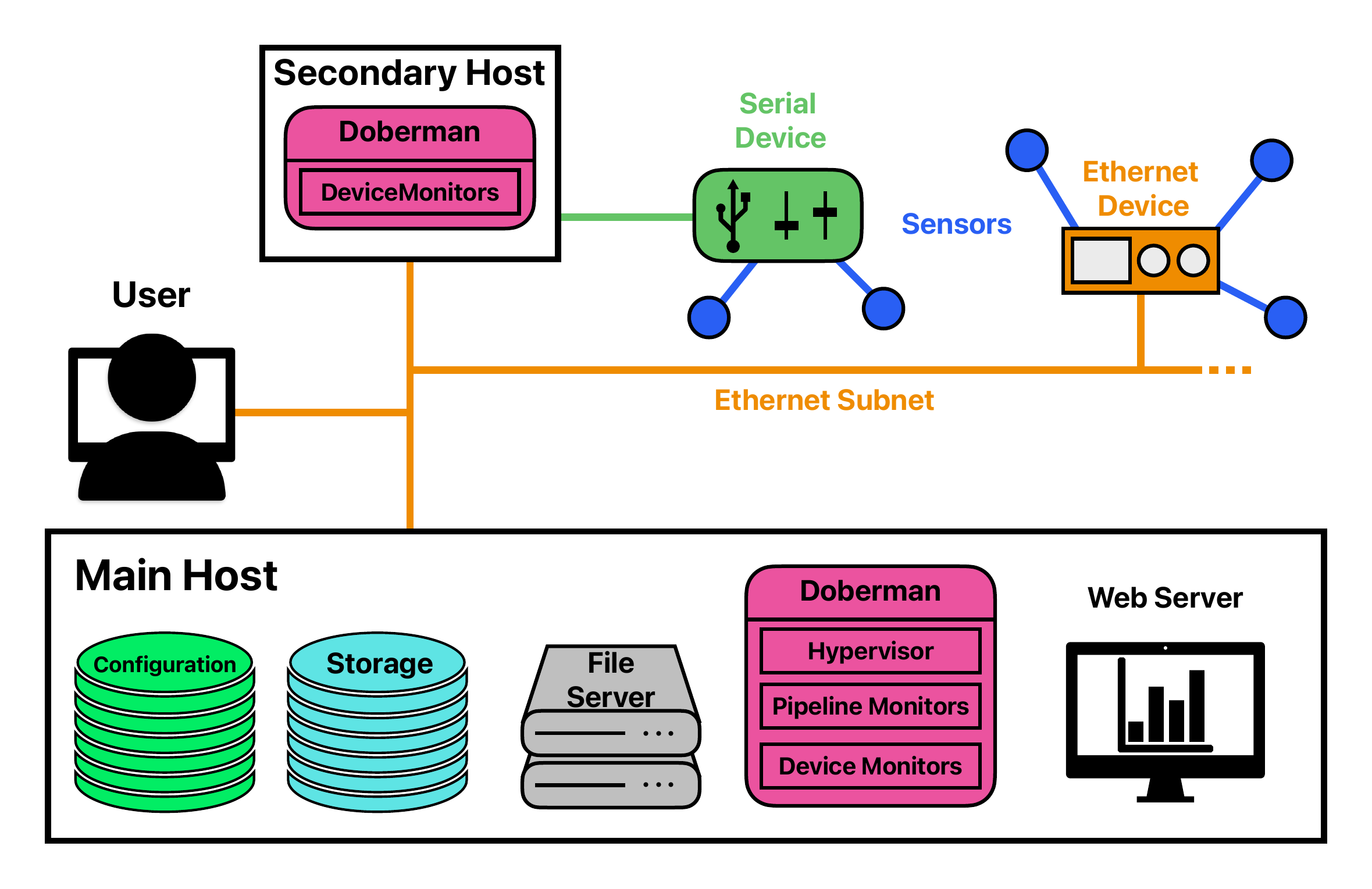}
    \caption{Overview of a distributed Doberman setup. A main server hosts the databases and, optionally, a network-mounted file system. The \texttt{Hypervisor} (central coordination) and \texttt{PipelineMonitors} (automation) typically run on the main server, while \texttt{DeviceMonitors} (interface to measuring devices) can be distributed across additional hosts to reduce cable lengths and improve resilience.}
    \label{fig:doberman_hardware_overview}
\end{figure}

\subsection{Deployment model}
\label{sec:deployment_model}

In a minimal configuration, Doberman can run on a single Linux host (e.g.\ an industrial mini-PC) that
connects to devices via Ethernet or a serial interface such as USB.
In this context, a \emph{device} denotes a physical instrument that provides access to one or more
measured or controlled quantities, while a \emph{sensor} represents a single such quantity
(e.g.\ a temperature, pressure, or digital state) handled by the slow control system.

In such a minimal setup, the configuration and time-series databases are typically hosted externally via managed database services, while the local host runs the central coordination and device readout processes.

A more typical laboratory deployment, in which Doberman operates as a distributed system with its own
on-premise database instances, is shown in \autoref{fig:doberman_hardware_overview}.
Here, a central server hosts the databases and the web interface and usually runs the
\texttt{Hypervisor} (\autoref{sec:hypervisor}), which coordinates communication and supervises system
components, as well as the \texttt{Pipeline Monitors} responsible for automation and alarm handling
(\autoref{sec:pipelines}).
The direct readout of devices is performed by dedicated \texttt{Device Monitors}
(\autoref{sec:sensors_and_devices}), which may run on the main server, or on one or more secondary hosts placed close to the experimental hardware and connected to the same Ethernet network.

Distributing device readout across multiple hosts improves robustness and operational flexibility: a failure of a single host affects only the subset of sensors connected to that machine, while the remaining system continues to operate normally. 
Affected components can be restarted on the same or on alternative hosts without interrupting global monitoring and alarm handling. 
In multi-host installations, a network-mounted software directory can be used to ensure consistent software versions across all machines and to simplify updates and maintenance.

Doberman is typically deployed within a closed and access-controlled laboratory network. Communication between components is therefore not encrypted by default and relies on the security of the underlying infrastructure.
\subsection{Databases}
\label{sec:databases}

Doberman employs a dual-database architecture to clearly separate configuration information from time-series measurement data, allowing each database to be optimized for its specific access patterns and performance requirements.

\paragraph{Configuration database (MongoDB).}
All configuration settings are stored in a MongoDB database \cite{mongodb}. A single MongoDB instance can host multiple experiments (separate databases), each organized into collections containing JSON-like documents. This structure provides flexibility because documents in a collection need not share identical fields, which simplifies schema evolution as new device types or features are added. In addition to configuration, Doberman stores log messages (above debug level) in MongoDB to enable centralized inspection via the web interface.

\paragraph{Time-series database (InfluxDB).}
Slow control measurements are stored in an InfluxDB OSS v2 instance \cite{influxdb}, which is optimized for time-series data. Similarly, a single InfluxDB instance can serve multiple experiments. Besides physics-relevant slow control quantities, Doberman can also record system-health metrics (e.g.\ CPU temperatures, load averages) to detect host-level issues that may affect reliability.

Data retention can be configured directly in InfluxDB. For current deployments, system monitoring data are stored for 30\,days, while slow control data are retained indefinitely, as storage requirements are modest. 
For instance, in the PANCAKE experiment described in \autoref{sec:deploy_pancake}, more than one year of total operation with about 100 slow control values written per second resulted in less than \SI{20}{GB} of stored data. For substantially larger deployments, InfluxDB provides built-in mechanisms to limit storage growth, for example by applying finite retention policies or by storing downsampled representations of older data.

\subsection{\texttt{Monitor} class}
\label{sec:monitor}

The fundamental building block of Doberman is the \texttt{Monitor} class. A \texttt{Monitor} manages multiple threads that run independently: each thread executes a task (e.g.\ reading out a sensor) and then sleeps for a configurable interval. This thread-based structure is used throughout Doberman for sensor readout, automation, supervision, and alarm distribution.

\begin{figure}[ht]
    \centering
    \includegraphics[width=\textwidth]{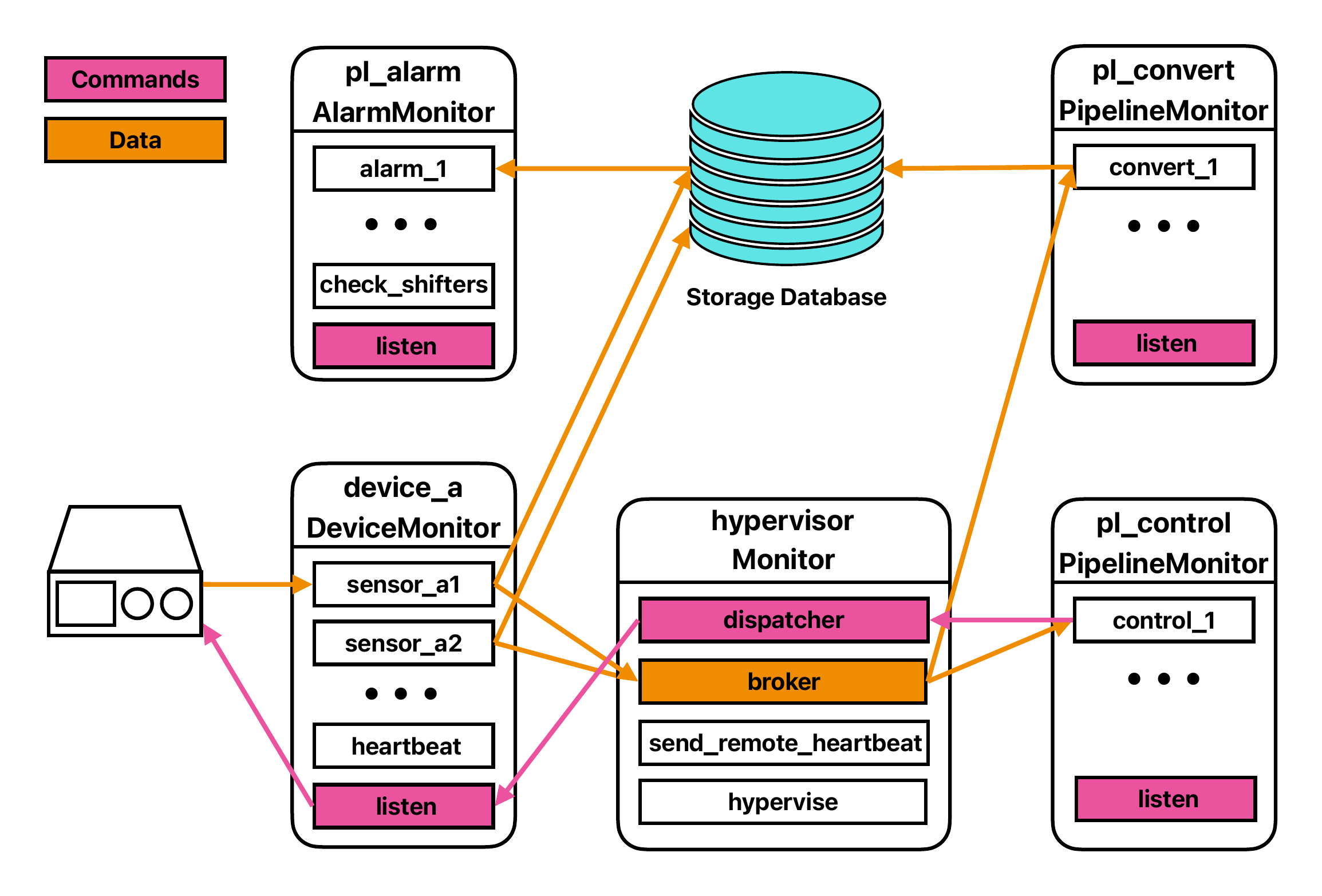}
    \caption{Overview of communication between Doberman Monitors and the storage database. Orange arrows indicate data flow and pink arrows command flow. Each box inside a Monitor represents a thread. All Monitors additionally access the configuration database (not shown).}
    \label{fig:monitor_overview}
\end{figure}

All major components of Doberman are implemented as specialized subclasses of the \texttt{Monitor} class. These include the central coordination unit (\texttt{Hypervisor}); \texttt{DeviceMonitors} interfacing with physical instruments; \texttt{PipelineMonitors} executing automated control and data-processing logic; and the \texttt{AlarmMonitor}, which evaluates alarm conditions and distributes notifications. The roles and interactions of these components are illustrated in \autoref{fig:monitor_overview} and are described in detail in the following sections.

\subsection{Hypervisor}
\label{sec:hypervisor}

While distributing the Doberman system over multiple hosts can be beneficial, operational experience shows that a central coordinating process is highly valuable. This role is served by the \texttt{Hypervisor}, which supervises component health, routes commands, and mediates data distribution. It has the following tasks:

\paragraph{Supervision and process management.}
The \texttt{Hypervisor} maintains a list of managed \texttt{Monitors} including the \texttt{PipelineMonitors} and a configurable set of \texttt{DeviceMonitors}. It checks responsiveness of all these \texttt{Monitors} via a periodic ping-pong message exchange and via device heartbeats recorded in the configuration database. If a \texttt{Monitor} becomes unresponsive, the \texttt{Hypervisor} attempts a restart. \texttt{Monitors} can be started locally or remotely: for remote hosts, the \texttt{Hypervisor} connects via SSH and starts the process in a terminal multiplexer session (e.g.\ \texttt{screen}).

\paragraph{Command distribution.}
Inter-component commands are transmitted via ZeroMQ \cite{zeromq} using a publish--subscribe pattern: the \texttt{Hypervisor} broadcasts messages, and each \texttt{Monitor} receives and filters commands addressed to it. Each command is associated with a unique identifier and is acknowledged by the receiving \texttt{Monitor} after successful execution. This acknowledgment mechanism allows the \texttt{Hypervisor} to track command delivery and execution, and log communication or process-level issues.

\paragraph{Data broker.}
In addition to command routing, the \texttt{Hypervisor} runs a data broker that distributes fresh sensor values to multiple consumers (e.g.\ pipelines) without requiring repeated database queries. As shown in \autoref{fig:monitor_overview}, pipelines can source data either from the broker (preferred for efficiency and reduced database load) or from the time-series database.

\paragraph{External heartbeat.}
To protect against rare but critical failure modes (e.g.\ power outage of the main server or network failures), Doberman supports an external heartbeat mechanism: the \texttt{Hypervisor} periodically updates a file on a remote system (in a different building, institution etc.). A lightweight watcher process on the remote system issues alarms if the heartbeat is not updated within a defined interval, preventing ``silent'' failures of the alarm infrastructure.

In the event of a temporary failure of the \texttt{Hypervisor} itself, running \texttt{Monitors} continue operating independently, including ongoing device readout and data storage. After the \texttt{Hypervisor} is restarted, \texttt{Monitors} automatically resume coordinated operation. Since measurement data are written directly to the time-series database, no data are lost during such interruptions.

\subsection{Sensors, devices, and plugins}
\label{sec:sensors_and_devices}

Integrating a new instrument requires configuration documents for the device and its sensors, and a device-specific Python \emph{plugin}. Database documents capture device settings and configurable parameters, while plugins encode the instrument-specific communication logic. Depending on the device, this logic can range from simple readout instructions to more complex implementations that include command execution and control.

\paragraph{Configuration documents.}
Device documents define connection parameters (e.g., IP address and port for Ethernet devices), the host on which the corresponding \texttt{DeviceMonitor} should run, and the list of sensors provided by the device. Sensor documents contain metadata used for organization and visualization, readout parameters (e.g., readout command and interval), and alarm configuration. 

Doberman further supports simple value transformations at the sensor level to convert raw device outputs into physically meaningful quantities. 
A typical example is a pressure sensor that provides a current output in the range 4--20\,mA, which can be converted into a pressure according to the calibration specified by the manufacturer. 

\paragraph{Plugins.}
Plugins are Python classes inheriting from \texttt{Device} or from interface-specific subclasses such as \texttt{LANDevice} or \texttt{SerialDevice}. For standard Ethernet and serial devices, connection handling and message transport are provided by the base classes. The plugin then specifies how to interpret returned messages, either via a simple pattern match (e.g.\ extracting a number with a regular expression) or via a custom parsing routine for more complex data formats.

\paragraph{Command execution and control.}
In addition to readout, Doberman can control devices (e.g., changing a PID setpoint, toggling a valve). Commands enter the system via the \texttt{Hypervisor} as JSON-like messages (originating from the web interface or from control pipelines) and are placed into a time-sorted queue for immediate or scheduled execution. The \texttt{Hypervisor} then broadcasts a hashed command to the appropriate \texttt{DeviceMonitor}. The \texttt{DeviceMonitor} translates the generic Doberman command syntax into the device-specific protocol using the plugin's \texttt{execute\_command} method and returns an acknowledgment containing the command hash.

\subsection{Pipelines}
\label{sec:pipelines}

Pipelines implement continuous processing and automation on top of sensor data. They are used to detect alarm conditions, compute derived quantities, and execute control logic (e.g.\ actuating valves). For usability, Doberman starts three \texttt{PipelineMonitors} corresponding to the different pipeline types: \texttt{alarm}, \texttt{convert}, and \texttt{control}.

Each pipeline is executed by its corresponding \texttt{PipelineMonitor} in a dedicated thread. An \texttt{AlarmMonitor} extends the generic \texttt{PipelineMonitor} with alarm distribution functionality (see \autoref{sec:alarm_distribution}). Pipelines can be \emph{inactive}, \emph{active}, or \emph{silent}. In silent mode, pipelines are evaluated normally but suppress command emission, alarm logging, and database writes, which is useful during commissioning, testing, or maintenance. Alarm pipelines additionally implement an auto-silence mechanism after issuing a notification to prevent message flooding.

Conceptually, pipelines are represented as directed graphs of nodes. Source nodes ingest data either from the live broker stream or from the database; intermediate nodes perform operations such as filtering, threshold checks, derivatives, or arbitrary expressions; and sink nodes write derived quantities to the database or trigger control actions. Some node types maintain internal buffers that store a configurable number of recent input values. These buffers enable operations that require temporal context, such as checking whether a threshold violation persists for several consecutive readouts, computing derivatives or averages, or suppressing short-lived fluctuations. For pipelines that depend on multiple input streams, \texttt{MergeNodes} combine incoming values into time-aligned payloads before forwarding them downstream. 

Pipeline behavior during temporary communication losses depends on the configured source nodes and control logic. If fresh broker data are unavailable, affected source nodes raise an error and the corresponding pipeline does not complete the affected cycle. Alarm pipelines can explicitly detect stale inputs using dedicated nodes that compare the timestamp of the last received value to the configured readout interval and issue an alarm if no new value has arrived within the allowed delay. For control pipelines, nodes may define default outputs that are applied when an error occurs, allowing safety-critical control logic to fall into a conservative state, for example closing a valve rather than leaving it open.

The latency of control pipelines is small compared to typical slow control readout intervals. In measurements with broker-based source nodes, the delay between publication of a threshold-crossing sensor value and the start of the corresponding command execution is below \SI{25}{ms}. For database-backed pipelines, where input values are queried from InfluxDB, the corresponding execution time is about \SI{100}{ms}. These values describe the internal Doberman processing time up to the start of command execution; the subsequent response time of the controlled hardware depends on the specific device. For typical slow control applications, such as temperature regulation, pressure control, or liquid-level handling, these latencies are negligible compared to the timescales of the controlled system.

\begin{figure}[ht]
    \centering
    \includegraphics[width=\textwidth]{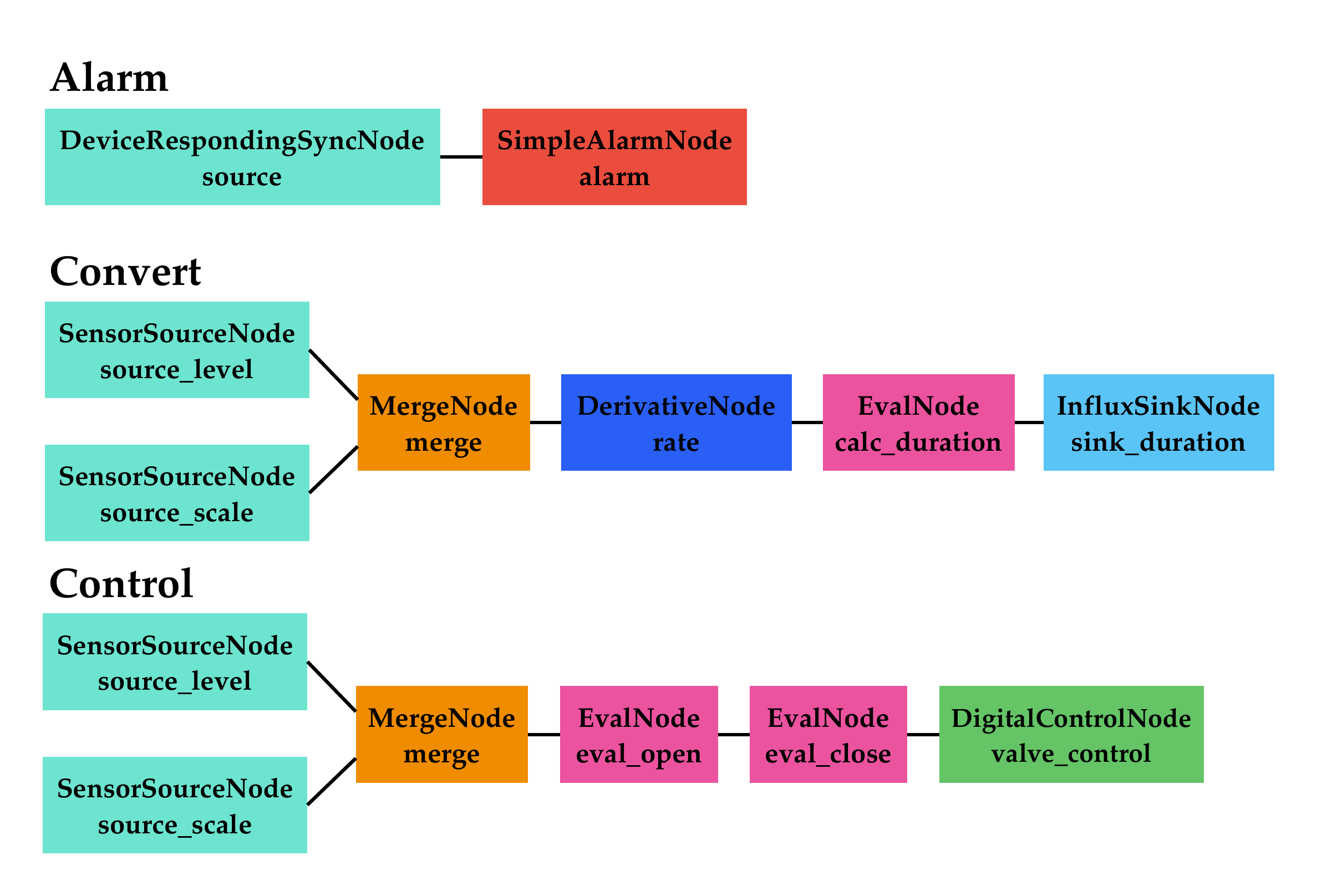}
    \caption{Examples of Doberman pipelines: a simple threshold alarm (top), a conversion pipeline computing a derived quantity (center), and a control pipeline actuating a valve based on sensor inputs (bottom).}
    \label{fig:pipeline_example}
\end{figure}

The three examples in \autoref{fig:pipeline_example} illustrate the typical use cases of Doberman pipelines. 
The alarm pipeline shown at the top implements a simple threshold check on a sensor value and issues an alarm when the device is unresponsive, or the measurement exceeds its configured limits. 
The conversion pipeline in the center combines multiple sensor inputs to compute a derived quantity. In this example, the liquid-nitrogen level measurement is used to determine the current consumption rate, which is then combined with the remaining mass measured by a scale to estimate the remaining supply duration of the nitrogen reservoir.
The control pipeline at the bottom uses the same sensor inputs to implement a simple control logic. The solenoid valve is opened when the liquid-nitrogen level in the receiving vessel falls below a configurable threshold and sufficient nitrogen remains in the supply dewar, and is closed once an upper fill level is reached or the supply dewar is depleted.

\subsection{Alarm distribution}
\label{sec:alarm_distribution}

Alarms are generated by \texttt{AlarmNodes} inside alarm pipelines and distributed by the \texttt{AlarmMonitor}. Doberman uses a leveled alarm concept: each alarm has a configured base level (from the sensor configuration) and an escalation level that increases if repeated alarm messages are sent without the alarm being cleared. Level-dependent parameters such as the auto-silence duration and the number of messages before escalation are defined centrally in the experiment configuration.

Recipients are grouped (e.g.\ \emph{shifters} for routine operation and \emph{experts} for severe conditions) based on contact documents in the configuration database. Depending on the effective alarm level, the \texttt{AlarmMonitor} sends notifications through one or more communication channels. In the current implementation, Doberman supports sending of e-mails, SMS via the SMS Creator by Nextp GmbH~\cite{smscreator}, and automated phone calls using the Voice API by Twilio Inc.~\cite{twilio}. The alarm framework is structured such that alternative notification backends can be integrated with minimal changes.

\section{Graphical user interface}
\label{sec:doberview}

Doberview is a web-based graphical user interface developed as the primary user-facing component of Doberman. It provides live visualization of slow control data, access to historical measurements, and interactive tools to configure sensors, pipelines, and alarms. Doberview is implemented as a Node.js application and uses the Bootstrap front-end framework to provide a responsive interface suitable for desktop and mobile devices.

The web application connects directly to the configuration database and the time-series backend, allowing users to inspect and modify system settings without manual database interaction.
Authentication can optionally be enabled via GitHub OAuth, granting write access to members of a designated GitHub organization while allowing unrestricted read-only access in less restrictive deployments. This approach has proven to be low-maintenance, as most collaborations already maintain a GitHub organization for code sharing. This system may need an update for experiments with more stringent security requirements, for example by restricting read access or by introducing user groups with different permission levels.

\subsection{Interactive overview}

\begin{figure}[htbp]
    \centering
    \includegraphics[width=\textwidth]{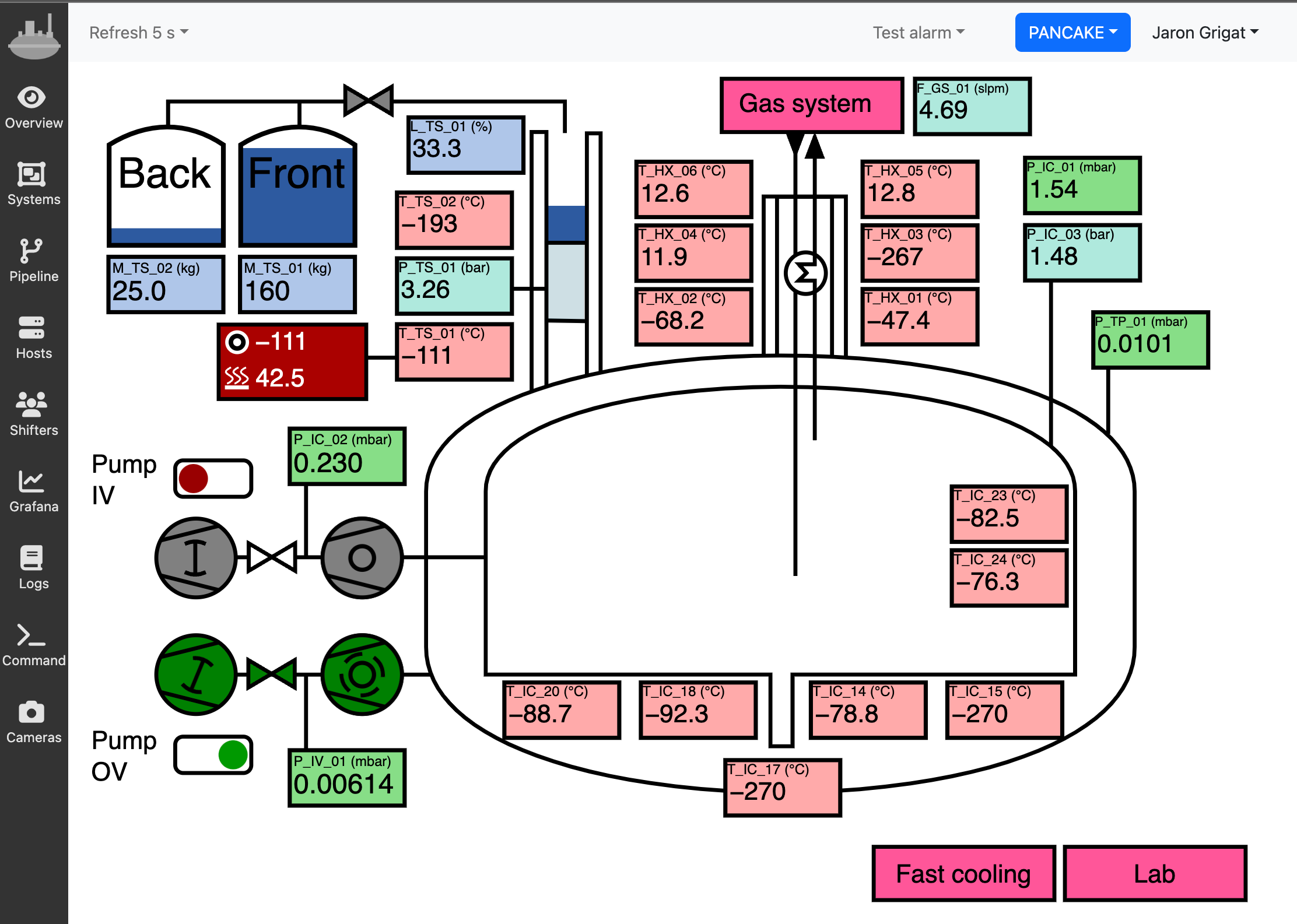}
    \caption{Interactive overview in Doberview for the PANCAKE experiment~\cite{Brown:2023vgf}. Sensor boxes display current values, pumps and valves change color or animate according to their state, and liquid levels are visualized based on derived quantities.}
    \label{fig:doberview_pid}
\end{figure}

If enabled, the overview page serves as the main landing page of Doberview. It is based on a user-defined scalable vector graphic (SVG) that visually represents the experimental setup, typically in the form of a simplified P\&ID diagram. Individual SVG elements are linked to sensors and devices, turning the static drawing into an interactive monitoring interface. An example of such an overview, showing the PANCAKE experiment described in \autoref{sec:deploy_pancake}, is shown in \autoref{fig:doberview_pid}.

Sensor elements display the most recent measurement values and update automatically. Binary states such as valve positions or pump status are represented through color changes and animations, while continuous quantities such as liquid-nitrogen fill levels are rendered as proportional graphical elements. Clicking on any interactive element opens the corresponding sensor detail view.

\subsection{Sensor inspection and control}

\begin{figure}[htbp]
    \centering
    \includegraphics[width=\textwidth]{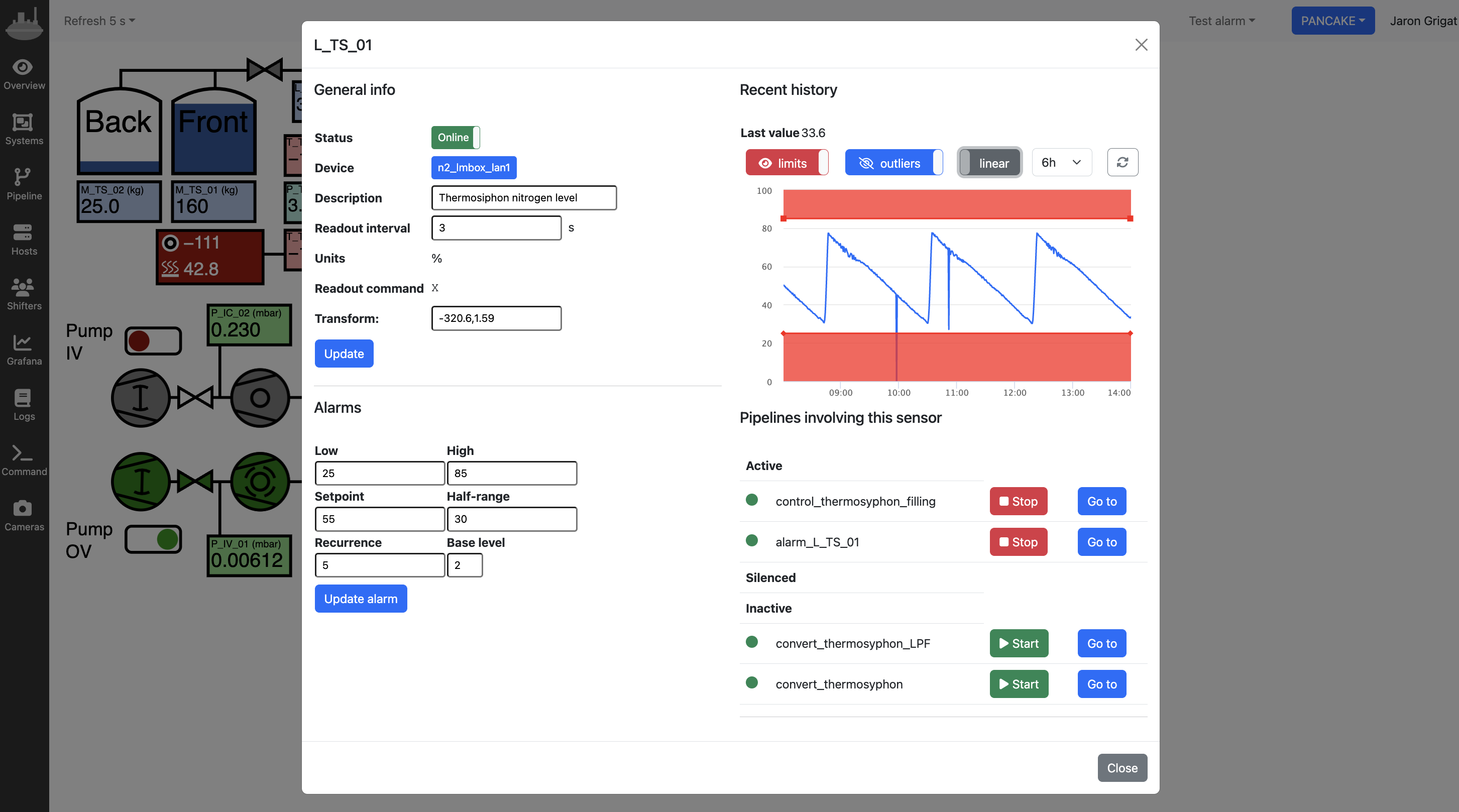}
    \caption{Example of a sensor detail window in Doberview. Configuration parameters and alarm settings are shown on the left, while recent measurement history and associated pipelines are displayed on the right.}
    \label{fig:doberview_sensor_modal}
\end{figure}

The sensor detail view provides a consolidated interface for monitoring, configuration, and control of individual sensors. An example of such a sensor detail window is shown in \autoref{fig:doberview_sensor_modal}. From this view, users can inspect recent measurement histories over selectable time ranges, adjust alarm thresholds, and interact with pipelines that rely on the sensor's data. If a sensor is associated with a controllable quantity (e.g.\ a valve state or a PID setpoint), corresponding control elements are displayed, allowing commands to be issued directly through the web interface.

\subsection{Pipeline management}

\begin{figure}[htbp]
    \centering
    \includegraphics[width=\textwidth]{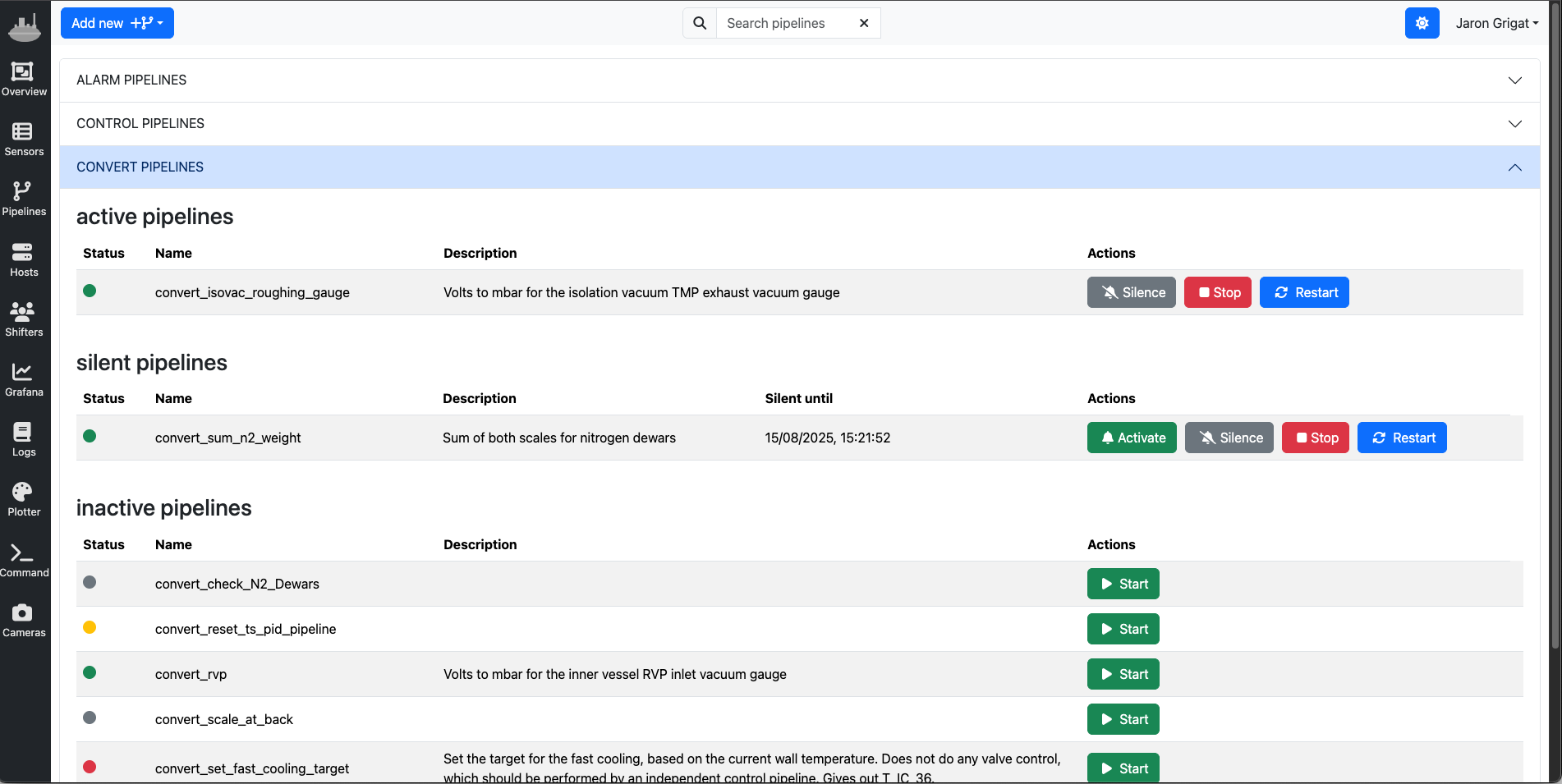}
    \caption{Pipeline overview in Doberview. Pipelines are grouped by type and execution state, with controls to start, stop, silence, or restart individual pipelines.}
    \label{fig:doberview_pipelines}
\end{figure}

Doberview provides a dedicated overview for managing pipelines, shown in \autoref{fig:doberview_pipelines}. 
Pipelines are grouped by type (alarm, convert, control) and by execution state (active, silent, or inactive). 
A color-coded status indicator provides immediate feedback on the most recent execution cycle: green indicates normal operation, red signals an error during execution, and yellow denotes transient startup conditions, for example while pipeline nodes with internal buffers are being initialized. 
Hovering over the indicator reveals additional information, such as the time of the last execution or the most recent error.

From this overview, users can start, stop, restart, or silence pipelines directly. 
Clicking on a pipeline opens a detailed configuration view in which the pipeline definition can be edited and visualized as a node graph. 
The same interface is used to create new pipelines from predefined templates, allowing users to define alarm, conversion, or control logic without manual interaction with the configuration database.

\subsection{Additional visualization tools}

In addition to the interactive overview and sensor-level inspection, Doberview provides complementary visualization tools for more advanced data exploration. These include a built-in plotting interface for rapid, ad hoc analysis as well as the option to embed external dashboards created with Grafana~\cite{grafana}.

\begin{figure}[htbp]
    \centering
    \includegraphics[width=\textwidth]{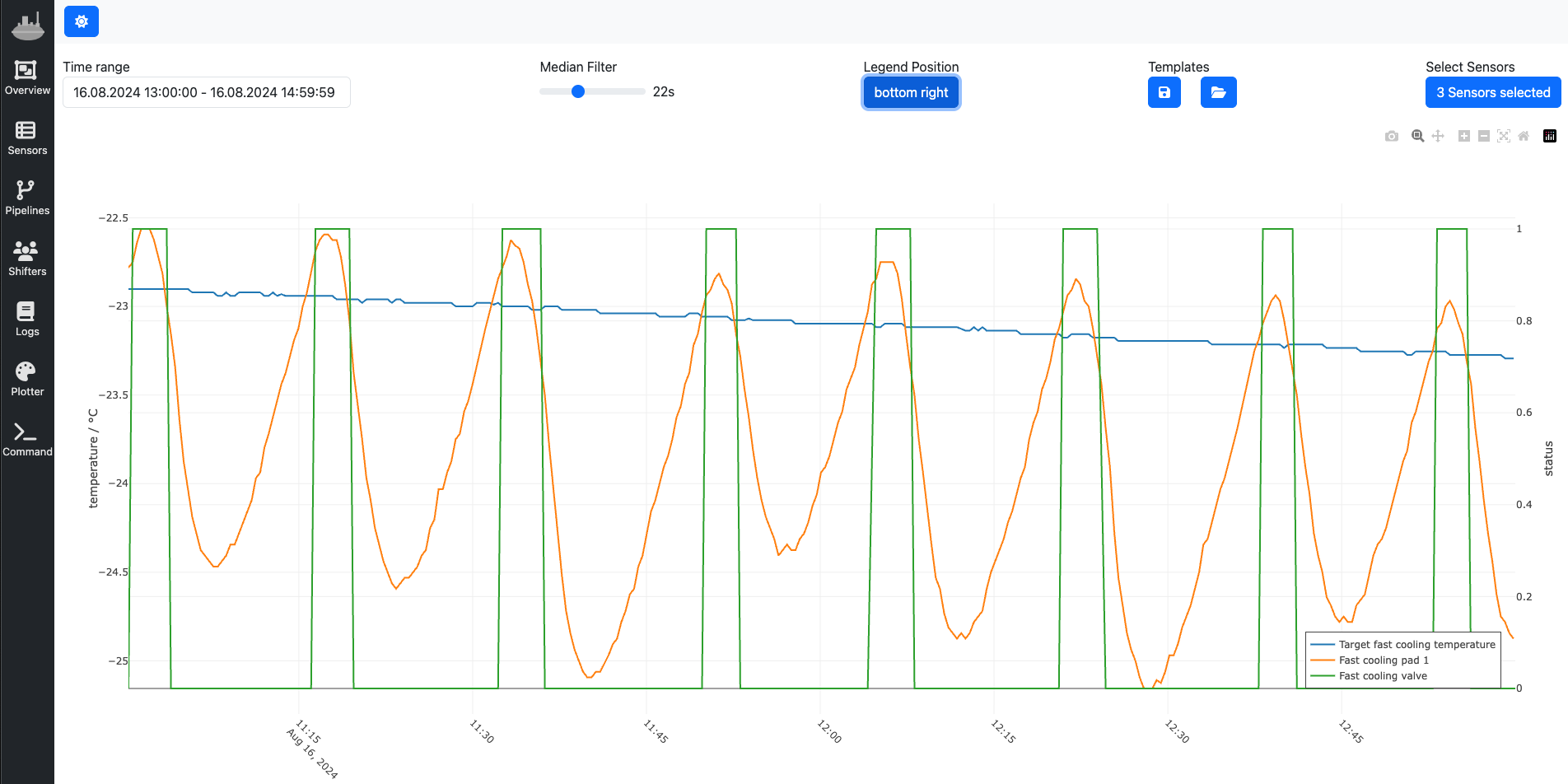}
    \caption{Plotter tab in Doberview showing a user-defined graph for the nitrogen-based cooling system of the PANCAKE experiment. Displayed are the target temperature (blue) from a convert pipeline, the measured temperature in the cooling circuit (orange), and the control signal opening the liquid-nitrogen valve (green).}
    \label{fig:doberview_plotter}
\end{figure}

The built-in \emph{Plotter} allows users to create custom time-series diagrams for arbitrary time intervals without additional configuration. An example is shown in \autoref{fig:doberview_plotter} for a nitrogen-based cooling system used to cool down the PANCAKE cryostat. In this case, the target temperature is provided by a conversion pipeline, the measured temperature is read by a sensor near a cooling pad, and the resulting control action, i.e. the opening of a solenoid valve, is displayed alongside both quantities.

Frequently used sensor combinations can be stored as templates in the configuration database and quickly reloaded. 
Together, the Doberview-native visualizations and embedded dashboards provide complementary tools optimized for different operational tasks. The interactive overview emphasizes the current state of the experiment with a clear visual grouping of related components, while sensor detail views allow rapid inspection of recent histories for individual measurements. The built-in plotter facilitates quick exploratory analyses and correlation checks across multiple sensors, whereas embedded Grafana dashboards are well suited for simultaneously monitoring larger sets of measurements over both short and extended time scales.

\section{Current deployments}
\label{sec:deployments}

Doberman has been used in several experimental setups operated by the Astroparticle Physics group at the University of Freiburg. The deployments span a small, remotely operated material screening facility in an underground laboratory, a small-scale liquid xenon R\&D platform, and a large and highly instrumented surface-level liquid xenon test facility. (It was also used for testbeam measurements of liquid scintillator-filled detector prototypes at CERN to develop the SHiP experiment~\cite{Brignoli:2025spc}, and to operate a sensitive radon emanation detector~\cite{Wiebe:2023kaa};  both projects are not further described here.) Together, these installations demonstrate that the same software stack supports very different requirements, ranging from low-rate monitoring on a single host to distributed, multi-host operation with extensive automation.

\subsection{GeMSE: Remote operation of a gamma-ray spectrometer}
\label{sec:deploy_gemse}

The Germanium Material and Meteorite Screening Experiment (GeMSE) is a low-background high-purity germanium (HPGe) gamma-ray spectrometer located in an underground laboratory in the Swiss Jura mountains \cite{gemse_og,gemse,meteorite}. Due to the remote location and the absence of permanent on-site personnel, reliable remote supervision and automated notification are essential.

For GeMSE, the complete Doberman stack runs on a single Linux host located in the underground laboratory and manages a compact set of critical parameters. In particular, Doberman supervises the bias voltage of the HPGe detector and the cryogenic cooling of the germanium crystal, which must be maintained at stable low temperatures throughout operation. Cooling is achieved using liquid nitrogen supplied from a large storage dewar, with an automated control pipeline periodically refilling a thermal bath connected to the detector. This setup enables fully autonomous operation over extended periods, such that manual replacement of the storage dewar is only required at intervals of several weeks. The use of Doberman allowed operating the spectrometer without unscheduled downtime over several years.

\subsection{XeBRA: A small detector test setup}
\label{sec:deploy_xebra}

The Xenon-Based Research Apparatus (XeBRA) is a liquid-xenon R\&D platform with target masses of order \SI{10}{kg}, used to study novel detector concepts for low-background xenon detectors~\cite{Dierle:2022zzh,Tonnies:2024vhy}. XeBRA comprises a vacuum-insulated cryostat, nitrogen-based xenon cooling, xenon purification and storage, and a dedicated photosensor DAQ \cite{xebra_instrumentation}. The slow control system monitors temperatures, pressures, flows, liquid xenon and liquid nitrogen levels, vacuum systems, and power supplies, and it provides automation for recurring operational tasks (e.g., nitrogen handling and cooldown-related procedures).

XeBRA shares its database infrastructure with the PANCAKE installation (see below) on a dedicated server in the laboratory. Device readout is performed by a set of \texttt{DeviceMonitors} running on hosts close to the setup, while the central services (databases, \texttt{Hypervisor}, pipelines, and web interface) run on the shared server. 
Since the XeBRA platform is used to host a variety of experimental configurations for testing novel detector concepts, the slow control system must accommodate frequent hardware changes and rapid integration of new instruments. Doberman's flexible plugin-based architecture allows new devices and sensors to be added with minimal effort, making it well suited for the iterative R\&D workflow typical of the XeBRA platform.

\subsection{PANCAKE: a complex, medium-scale experiment}
\label{sec:deploy_pancake}

PANCAKE is a large liquid-xenon test facility with a current xenon gas inventory of about 400\,kg~\cite{Brown:2023vgf}. Its inner diameter of about 3\,m allows testing large-scale detector components in cryogenic liquid xenon. It was recently used to demonstrate the operation of a liquid xenon TPC with active mass of about 130\,kg in an unshielded, above-ground environment~\cite{pancake_tpc}. Its slow control system, the most complex Doberman deployment to date, continuously reads out approximately 300 sensors across several dozen instruments and provides automated monitoring and control throughout multi-month campaigns, including cooldown, filling, and stable long-term operation. The requirement to operate such a setup with a small on-call team was a major motivation for developing Doberman's control-pipeline framework (\autoref{sec:pipelines}).

As in the XeBRA deployment, the configuration and time-series databases, the \texttt{Hypervisor}, the pipeline execution processes, and the Doberview web interface are hosted on the shared central server. 
The readout of most devices runs on additional industrial hosts deployed close to the respective subsystems, reducing cable lengths and improving overall robustness. These secondary hosts are implemented using Revolution~Pi modules manufactured by KUNBUS GmbH \cite{revpi}. The DIN-rail-mounted industrial PCs are based on the Raspberry~Pi Compute Module, run a Linux-based operating system, and comply with European industrial standards for programmable controllers. In the PANCAKE deployment, they have operated stably over extended periods, supporting continuous slow control operation. 

\begin{figure}[ht]
     \centering
     \includegraphics[width=0.75\textwidth]{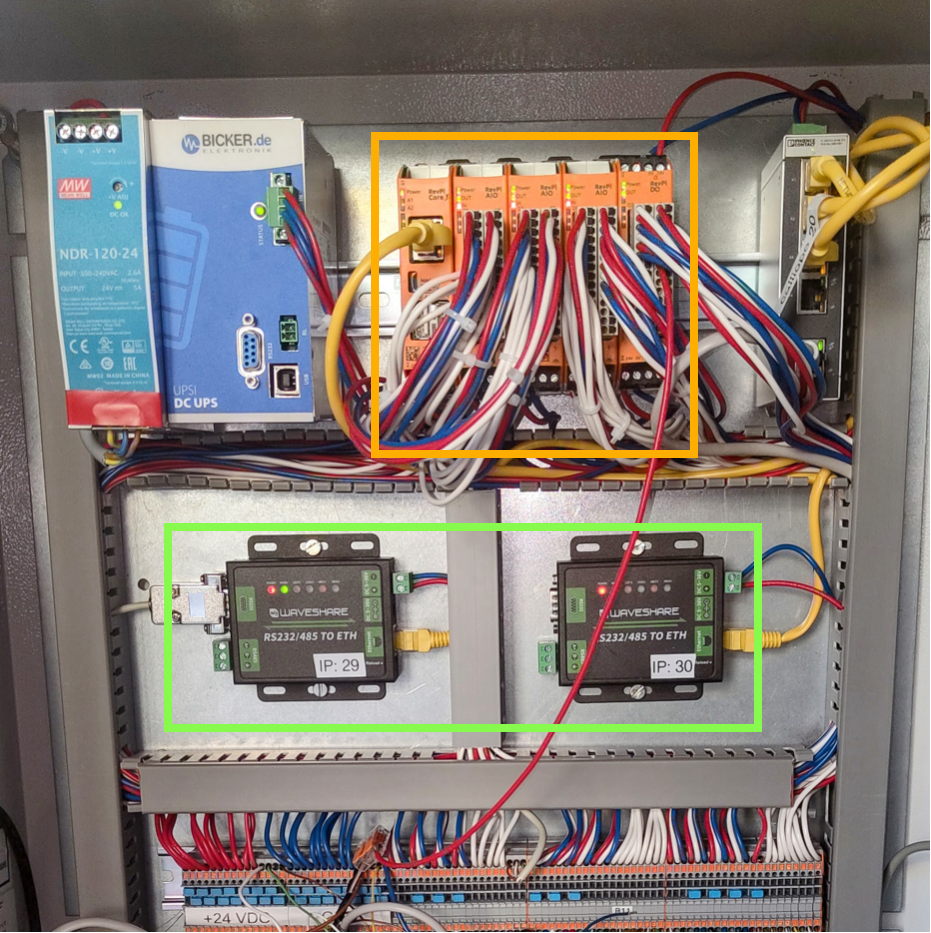}
     \caption{One of the PANCAKE slow control cabinets, containing a \SI{24}{VDC} power supply, an uninterruptible power supply (UPS), a Revolution~Pi core with modular analog and digital I/O modules (orange), and a rail-mounted Ethernet switch. Serial instruments are integrated via WaveShare serial-to-Ethernet converters (green). Sensor cables are routed through a terminal block at the bottom.}
     \label{fig:scbox}
\end{figure}

Each core module can be extended with multiple modular I/O units, allowing flexible integration of a large number of sensors and actuators. In the PANCAKE setup, the cores are equipped with analog I/O and digital output modules providing configurable voltage and current inputs (e.g., 0--5\,V or 4--20\,mA), RTD inputs for PT100/PT1000 temperature sensors, and digital outputs for controlling actuators such as solenoid valves. An example installation inside a control cabinet is shown in \autoref{fig:scbox} (orange frame).

Most serial devices in PANCAKE are integrated using WaveShare serial-to-Ethernet converters \cite{waveshare}, two of which are shown in \autoref{fig:scbox} (green frame). These modules provide both RS-232 and RS-485 interfaces, enabling reliable readout of a wide range of legacy and industrial devices over the laboratory network. 

A large fraction of the PANCAKE slow control channels correspond to temperature measurements distributed throughout the cryostat and its supporting systems. For this purpose, the Web Thermometer~8x manufactured by Wiesemann \& Theis GmbH has proven to be a reliable solution \cite{wut}. Each device supports eight PT100 temperature channels and provides network-based access via an HTTP interface, allowing straightforward integration through a dedicated Doberman plugin. The corresponding plugins, along with those for other PANCAKE-specific devices, are available in the public repository \cite{doberman_pancake}.

Doberman was essential for the stable and safe operation of the PANCAKE detector during two extended, multi-month run campaigns, including the most recent campaign in late 2024. Continuous monitoring and automated control of key parameters such as xenon pressure, cryostat temperatures, and liquid levels enabled reliable long-term operation of a  Time Projection Chamber (TPC) and its cryogenic infrastructure. 
The alarm system and graphical user interfaces provided rapid feedback on detector conditions and facilitated timely intervention when required. The resulting stability of cryogenic and high-voltage parameters ensured well-defined and reproducible operating conditions during approximately sixty days of TPC operation, forming the basis for the data analysis presented in \cite{pancake_tpc}.

\subsection{Operational performance and scaling}
\label{sec:performance_scaling}

The long-term operation of the deployments described above demonstrates that Doberman has modest hardware requirements. As an example, the PANCAKE slow control system runs mainly on a server equipped with an Intel Xeon E3-1231 v3 CPU with four physical cores/eight logical threads. Each \texttt{Monitor} typically requires about \SI{50}{MB} of resident memory. During normal operation, the \texttt{Hypervisor} and \texttt{PipelineMonitors} each use less than \SI{1}{\percent} CPU on this system, while representative \texttt{DeviceMonitors} remain below \SI{0.5}{\percent}.

To probe scalability beyond the current deployments, additional software-only test devices were operated on the same system. Ten such devices with 100~sensors each were read out at \SI{1}{\hertz} per sensor over several hours, resulting in an aggregate configured measurement rate of \SI{1000}{\hertz}. This is about an order of magnitude larger than the configured aggregate measurement rate of the PANCAKE deployment described in \autoref{sec:deploy_pancake}. Each synthetic \texttt{DeviceMonitor} required about \SI{50}{MB} of resident memory and about \SI{20}{\percent} CPU. The increased readout rate had no significant impact on the CPU usage of the \texttt{Hypervisor}, indicating that the dominant load is localized in the \texttt{DeviceMonitor} processes. Since \texttt{DeviceMonitors} can be distributed over multiple hosts, this load can be scaled horizontally.

A second test used one software-only test device together with 100~active control pipelines. Each pipeline read one sensor value and issued a binary control command, resulting in an aggregate command rate of about \SI{100}{\hertz}. In this configuration, the \texttt{Hypervisor} used about \SI{6}{\percent} CPU, while the control \texttt{PipelineMonitor} used about \SI{21}{\percent} CPU. These values remain moderate on the comparatively modest server hardware used in the test.

Together, these measurements show that the central Doberman processes do not form a bottleneck at rates well above those required by the largest current deployment. The main readout load resides in the \texttt{DeviceMonitors}, which can be distributed across hosts, while the database backend based on InfluxDB is designed for substantially higher write rates than those encountered in typical slow-control applications. These software-only tests therefore support the use of Doberman for medium-scale experiments beyond the currently operated deployments. A full validation of network-level scaling and distributed failure modes, however, will require dedicated tests on a physically distributed multi-host installation.

\section{Conclusions and outlook}
\label{sec:conclusion}

Doberman is a lightweight, modular, and open-source slow control system designed to bridge the gap between heavyweight and costly SCADA/EPICS-style frameworks and ad hoc laboratory scripts. Its plugin-based device integration, distributed \texttt{Monitor} architecture with a central \texttt{Hypervisor}, and node-based pipeline framework enable reliable sensor readout, automation, and transparent alarm handling across heterogeneous experimental hardware. The accompanying web interface, Doberview, provides an integrated environment for live monitoring, configuration, and control, supporting both routine operation and rapid response to exceptional conditions.

Doberman and Doberview have matured through multiple iterations and have been validated in day-to-day operation across several experiments at the University of Freiburg, ranging from a small remote underground facility to a large, highly instrumented liquid xenon test platform. Both projects are released under permissive open-source licenses, with source code, documentation, and a growing collection of device plugins available online \cite{doberman_github,doberview,doberman_pancake,doberman_xebra}. Interest from external groups indicates that the approach is broadly applicable to laboratory-scale and medium-scale experimental infrastructures. 
Experience from the current deployments, together with the performance tests presented above, indicates that Doberman can be used in experiments beyond the scale of the systems operated so far, while network-level scaling and distributed failure modes remain deployment-dependent aspects that should be evaluated for substantially larger installations. 
Doberman is foreseen to be used as the slow control system for the upcoming light dark matter experiment DELight~\cite{delight}, which may motivate targeted refinements where required by specific experimental needs, although the system is considered functionally mature and no major architectural changes are foreseen.

\subsection*{Acknowledgments}

We would like to thank our colleagues using Doberman for their valuable feedback to improve the system. Over the years of development, this work was supported in part by the German Federal Ministry of Research, Technology and Space through the ErUM-Pro grant 05A23VF1, by the European Research Council (ERC) grant No. 724320 (ULTIMATE), by the German Research Foundation (DFG) through INST 39/1095-1 FUGG and the Research Training Group RTG2044, and the Structure and Innovation Fund of the state of Baden-W\"urttemberg (SI-BW).


 \bibliographystyle{JHEP}
 \bibliography{citations}

@article{delight,
  title = "{DELight: a Direct search Experiment for Light dark matter with superfluid helium}",
  shorttitle = {DELight},
  author = {von Krosigk, Belina and Eitel, Klaus and Enss, Christian and Ferber, Torben and Gastaldo, Loredana and Kahlhoefer, Felix and Kempf, Sebastian and Klute, Markus and Lindemann, Sebastian and Schumann, Marc and Toschi, Francesco and Valerius, Kathrin},
  year = {2023},
  month = jul,
  journal = {SciPost Physics Proceedings},
  number = {12},
  pages = {016},
  issn = {2666-4003},
  doi = {10.21468/SciPostPhysProc.12.016},
  eprint = {arXiv:2209.10950},
  archivePrefix = {arXiv},
  primaryClass = {hep-ex},
  urldate = {2025-11-15},
}

@misc{doberman_github,
  author       = {Masson, Darryl and Grigat, Jaron and Brown, Adam and Ram\'irez Garc\'ia, Diego and Luce, Tiffany and Zappa, Philipp},
  title        = {{AG-Schumann/Doberman}: v6.0.0},
  year         = {2026},
  publisher    = {Zenodo},
  version      = {v6.0.0},
  howpublished = {\url{https://doi.org/10.5281/zenodo.20043032}}
}

@misc{doberview,
  author       = {Grigat, Jaron and Brown, Adam and M\"uller, Julia and Masson, Darryl and Luce, Tiffany},
  title        = {{AG-Schumann/doberview2}: v1.0},
  year         = {2026},
  publisher    = {Zenodo},
  version      = {v1.0},
  howpublished = {\url{https://doi.org/10.5281/zenodo.20043125}}
}

@misc{doberman_pancake,
  author       = {Grigat, Jaron and Brown, Adam and Masson, Darryl and Luce, Tiffany},
  title        = {{AG-Schumann/doberman\_pancake}: v1.0},
  year         = {2026},
  publisher    = {Zenodo},
  version      = {v1.0},
  howpublished = {\url{https://doi.org/10.5281/zenodo.20054388}}
}

@misc{doberman_xebra,
  author       = {Grigat, Jaron and Masson, Darryl and Luce, Tiffany},
  title        = {{AG-Schumann/doberman\_xebra}: v1.0},
  year         = {2026},
  publisher    = {Zenodo},
  version      = {v1.0},
  howpublished = {\url{https://doi.org/10.5281/zenodo.20054466}}
}

@article{doberman_og,
  title = {A versatile and light-weight slow control system for small-scale applications},
  author = {Zappa, Philipp and B{\"u}tikofer, Lukas and Coderre, Daniel and Kaminsky, Basho and Schumann, Marc and von Sivers, Moritz},
  year = 2016,
  month = sep,
  journal = {Journal of Instrumentation},
  volume = {11},
  number = {09},
  eprint = {arXiv:1607.08189},
  primaryclass = {physics},
  pages = {T09003-T09003},
  issn = {1748-0221},
  doi = {10.1088/1748-0221/11/09/T09003},
  urldate = {2025-07-24},
  archiveprefix = {arXiv}
}

@misc{epics,
  author       = {{EPICS Collaboration}},
  title        = {{EPICS: Experimental Physics and Industrial Control System}},
  howpublished = {\url{https://epics-controls.org/}, accessed on May 6, 2026}
}

@article{epics2,
  title = {The experimental physics and industrial control system architecture: past, present, and future},
  shorttitle = {The experimental physics and industrial control system architecture},
  author = {Dalesio, Leo R. and Hill, Jeffrey O. and Kraimer, Martin and Lewis, Stephen and Murray, Douglas and Hunt, Stephan and Watson, William and Clausen, Matthias and Dalesio, John},
  year = 1994,
  month = dec,
  journal = {Nuclear Instruments and Methods in Physics Research Section A: Accelerators, Spectrometers, Detectors and Associated Equipment},
  volume = {352},
  number = {1},
  pages = {179--184},
  issn = {0168-9002},
  doi = {10.1016/0168-9002(94)91493-1},
  urldate = {2025-10-22}
}

@article{gemse,
  title = "{GeMSE: a Low-Background Facility for Gamma-Spectrometry at Moderate Rock Overburden}",
  shorttitle = {GeMSE},
  author = {Garc{\'i}a, Diego Ram{\'i}rez and Baur, Daniel and Grigat, Jaron and Hofmann, Beda A. and Lindemann, Sebastian and Masson, Darryl and Schumann, Marc and von Sivers, Moritz and Toschi, Francesco},
  year = 2022,
  month = apr,
  journal = {Journal of Instrumentation},
  volume = {17},
  number = {04},
  eprint = {arXiv:2202.06540},
  primaryclass = {physics},
  pages = {P04005},
  issn = {1748-0221},
  doi = {10.1088/1748-0221/17/04/P04005},
  urldate = {2025-08-21},
  archiveprefix = {arXiv}
}

@article{gemse_og,
  title = "{The GeMSE facility for low-background $\gamma$-ray spectrometry}",
  author = {von Sivers, M. and Hofmann, B.~A. and Ros{\'e}n, {\AA}.~V. and Schumann, M.},
  year = 2016,
  month = dec,
  journal = {Journal of Instrumentation},
  volume = {11},
  number = {12},
  eprint = {arXiv:1606.03983},
  primaryclass = {physics.ins-det},
  pages = {P12017},
  issn = {1748-0221},
  doi = {10.1088/1748-0221/11/12/P12017},
  urldate = {2025-08-21},
  archiveprefix = {arXiv},
  langid = {english}
}

@misc{grafana,
  author       = {{Grafana Labs}},
  title        = {{Grafana monitoring and visualization platform}},
  howpublished = {\url{https://grafana.com/}, accessed on May 6, 2026}
}

@misc{influxdb,
  author       = {{InfluxData, Inc.}},
  title        = {{InfluxDB}},
  howpublished = {\url{https://www.influxdata.com/}, accessed on May 6, 2026}
}

@misc{mongodb,
  author       = {{MongoDB, Inc.}},
  title        = {{MongoDB}},
  howpublished = {\url{https://www.mongodb.com/}, accessed on May 6, 2026}
}

@article{pancake_tpc,
  title        = {Operating a large-diameter dual-phase liquid xenon TPC in the unshielded PANCAKE facility},
  author       = {M{\"u}ller, Julia and Grigat, Jaron and Glade-Beucke, Robin and Lindemann, Sebastian and Luce, Tiffany and Madduri, Gnanesh Chandra and Reininghaus, Jens and Schumann, Marc and Softley-Brown, Adam and Stevens, Andrew},
  year         = {2026},
  eprint       = {arXiv:2601.15938},
  archiveprefix= {arXiv},
  primaryclass = {physics.ins-det},
  note         = {Submitted January 2026}
}

@article{Brown:2023vgf,
    author = "Brown, Adam and others",
    title = "{PANCAKE: a large-diameter cryogenic test platform with a flat floor for next generation multi-tonne liquid xenon detectors}",
    eprint = "arXiv:2312.14785",
    archivePrefix = "arXiv",
    primaryClass = "physics.ins-det",
    doi = "10.1088/1748-0221/19/05/P05018",
    journal = "Journal of Instrumentation",
    volume = "19",
    number = "05",
    pages = "P05018",
    year = "2024"
}

@misc{revpi,
  author       = {{KUNBUS GmbH}},
  title        = {{RevPi Core and RevPi Core 3 datasheet}},
  howpublished = {\url{https://revolutionpi.com/fileadmin/downloads/legacy/Datasheet_RevPi_Core_RevPi_Core3.pdf}, accessed on May 6, 2026}
}

@misc{smscreator,
  author       = {{Nextp GmbH}},
  title        = {{SMS Creator bulk messaging service}},
  howpublished = {\url{https://www.smscreator.de/}, accessed on May 6, 2026}
}

@misc{twilio,
  author       = {{Twilio Inc.}},
  title        = {{Twilio communications API}},
  howpublished = {\url{https://www.twilio.com/}, accessed on May 6, 2026}
}

@misc{waveshare,
  author       = {{Waveshare}},
  title        = {{Serial-to-Ethernet converter (RS-232/RS-485)}},
  howpublished = {\url{https://www.waveshare.com/product/iot-communication/wired-comm-converter/ethernet-to-uart-rs232-rs485.htm}, accessed on May 6, 2026}
}

@misc{wut,
  author       = {{Wiesemann \& Theis GmbH}},
  title        = {{Web Thermometer 8x model 57778 data sheet}},
  howpublished = {\url{https://www.wut.de/e-57778-ww-daus-000.php}, accessed on May 6, 2026}
}

@article{xebra_instrumentation,
  title = {The XeBRA platform for liquid xenon time projection chamber development},
  author = {Baur, Daniel and Bismark, Alexander and Brown, Adam and Dierle, Julia and Fischer, Horst and {Glade-Beucke}, Robin and Grigat, Jaron and Kaminsky, Basho and Kuger, Fabian and Lindemann, Sebastian and Masson, Darryl and Meinhardt, Patrick and Rajado Silva, Mariana and Schumann, Marc and T{\"o}nnies, Florian and Toschi, Francesco},
  year = 2023,
  month = feb,
  journal = {Journal of Instrumentation},
  volume = {18},
  number = {02},
  eprint = {arXiv:2208.14815},
  primaryclass = {physics.ins-det},
  pages = {T02004},
  publisher = {IOP Publishing},
  issn = {1748-0221},
  doi = {10.1088/1748-0221/18/02/T02004},
  urldate = {2025-08-26},
  archiveprefix = {arXiv},
  langid = {english}
}

@misc{zeromq,
  author       = {{The ZeroMQ Project}},
  title        = {{ZeroMQ}},
  howpublished = {\url{https://zeromq.org/}, accessed on May 6, 2026}
}

@article{Brignoli:2025spc,
    author = "Brignoli, A. and others",
    title = "{Performance of prototypes with different reflector materials for the SHiP liquid scintillator surrounding background tagger}",
    eprint = "arXiv:2503.10250",
    archivePrefix = "arXiv",
    primaryClass = "physics.ins-det",
    doi = "10.1088/1748-0221/20/07/P07023",
    journal = "Journal of Instrumentation",
    volume = "20",
    number = "07",
    pages = "P07023",
    year = "2025"
}

@article{Tonnies:2024vhy,
    author = {T{\"o}nnies, Florian and Brown, Adam and Kiyim, Baris and Kuger, Fabian and Lindemann, Sebastian and Meinhardt, Patrick and Schumann, Marc and Stevens, Andrew},
    title = "{Proportional scintillation in liquid xenon: demonstration in a single-phase liquid-only time projection chamber}",
    eprint = "arXiv:2405.10687",
    archivePrefix = "arXiv",
    primaryClass = "physics.ins-det",
    doi = "10.1088/1748-0221/19/09/P09032",
    journal = "Journal of Instrumentation",
    volume = "19",
    number = "09",
    pages = "P09032",
    year = "2024"
}

@article{Dierle:2022zzh,
    author = "Dierle, Julia and Brown, Adam and Fischer, Horst and Glade-Beucke, Robin and Grigat, Jaron and Kuger, Fabian and Lindemann, Sebastian and Silva, Mariana Rajado and Schumann, Marc",
    title = "{Reduction of $^{222}\hbox {Rn}$-induced backgrounds in a hermetic dual-phase xenon time projection chamber}",
    eprint = "arXiv:2209.00362",
    archivePrefix = "arXiv",
    primaryClass = "physics.ins-det",
    doi = "10.1140/epjc/s10052-022-11151-w",
    journal = "Eur. Phys. J. C",
    volume = "83",
    number = "1",
    pages = "9",
    year = "2023"
}

@article{meteorite,
author = {Ros{\'e}n, {\AA}.~V. and Hofmann, Beda A. and Preusser, Frank and Gnos, Edwin and Eggenberger, Urs and Schumann, Marc and Szidat, Sönke},
title = {Meteorite terrestrial ages in Oman based on gamma spectrometry and sediment dating, focusing on the Ramlat Fasad dense collection area},
journal = {Meteoritics \& Planetary Science},
volume = {56},
number = {11},
pages = {2017-2034},
doi = {https://doi.org/10.1111/maps.13758},
year = {2021}
}

@article{Wiebe:2023kaa,
    author = "Wiebe, Daniel and Lindemann, Sebastian and Schumann, Marc",
    title = "{A high-sensitivity radon emanation detector system for future low-background experiments}",
    eprint = "arXiv:2309.04514",
    archivePrefix = "arXiv",
    primaryClass = "physics.ins-det",
    doi = "10.1088/1748-0221/19/04/P04014",
    journal = "Journal of Instrumentation",
    volume = "19",
    number = "04",
    pages = "P04014",
    year = "2024"
}

\end{document}